\documentclass[twocolumn]{article}

\usepackage{scrextend}
\usepackage{lineno,hyperref}
\usepackage{threeparttable}
\usepackage{subcaption}
\usepackage{graphicx}
\usepackage{adjustbox}
\usepackage{tablefootnote}
\usepackage{amsmath}
\usepackage{wasysym}
\usepackage[affil-it]{authblk}

\begin{document}

\title{\vspace{-2.0cm}\textbf{WATER AND AIR CONSUMPTION ABOARD INTERSTELLAR ARKS}}
   
\author{Fr\'ed\'eric Marin\textsuperscript{1} \& Camille Beluffi\textsuperscript{2}\\
{\small 1 Universit\'e de Strasbourg, CNRS, Observatoire astronomique de Strasbourg, UMR 7550, F-67000 Strasbourg, France\\
2 CASC4DE, Le Lodge, 20, Avenue du Neuhof, 67100 Strasbourg, France}}
              
\date{Dated: \today}

\twocolumn[
  \begin{@twocolumnfalse}
    \maketitle
    \begin{abstract}
    The architecture of a large interstellar spaceship, which is capable of serving as a living environment for a
    population over many generations, is mainly dictated by the needs of said population in terms of food, water 
    and breathable gases. These resources cannot be stored for the entire duration of a journey that goes on for
    several centuries, so they must be produced \textit{in situ}. In order to determine the quantities necessary 
    for the survival of the population, it is imperative to precisely estimate their needs. In this article, we 
    focus on accurate simulations of the water and air (oxygen) requirements for any type of population in order 
    to be able to provide precise constraints on the overall architecture of an interstellar ark (the requirements 
    in terms of food having already been studied in a previous publication). We upgrade our agent-based, numerical,
    Monte Carlo code HERITAGE to include human physiological needs. Our simulations show that, for a crew of about
    1\,100 crew members (each characterized with individual anthropometric and biological data), 1.8 $\times$ 10$^8$ 
    litres of oxygen are annually required, together with 1.1 $\times$ 10$^6$ litres of water. Those results do not account 
    for oxygen and water used in growing plants, but they give us an insight of how much resources are needed in 
    the spaceship. We also review the best methods for generating water from waste gases (namely carbon dioxide and
    dihydrogen) and how such system could complement the oxygen-supplying biospheres inside multi-generational spaceship
    to form a closed and controlled environment.
    \end{abstract}
    
    \vspace{3\baselineskip}
  \end{@twocolumnfalse}
]

\section{Introduction}
\label{Introduction}
In order to prepare for a long-term mission in an environment where resources are scarce, it is necessary to plan and budget 
for the equipment to bring, to build on site and to recycle. If we take the real example of the French commercial ship 
\textit{Clairon et Reine}, a brig which regularly travelled from Marseille (France) to Smyrna (Turkey) between 1827 and 1836, 
we can study how the rationing in food was important and precise in order to carry out long sea crossings \cite{Reynaud1973}. 
According to the ship's documents found in 1973, each of the $\sim$ 10 sailors obtained a ration of nearly 4\,200 daily 
kilo-calories in bread, meat, fish, vegetables and cereals, which matches their intense physical activity level. Thousands of 
litres of liquids (mainly alcohol and water) were loaded on board before each departure, so that month-long trips could be 
achieved \cite{Reynaud1975}. This represents several tens of tonnes of food which tended to spoil after a few weeks. Now, 
let us compare this situation to space travels, where breathable air becomes a valuable resources that must be taken into 
account in the equation. In addition, it will not be possible to fish in the vacuum of space, so food cannot be easily 
replenished. Storing enough food, water and nutriments, together with a breathable mixture of gases, represents a challenge. 
First it will require physical space and add mass to the space shuttle or spaceship, which drives higher costs (larger ships, stronger
propulsion systems). Second, water and food will eventually spoil in time despite the best industrial methods for food preservation 
and storage \cite{Kokini1995,Subramaniam2000}. These problems are at the centre of contemporary thinking for the human exploration 
and colonization of Mars \cite{Nelson1993,Nelson1995,Alling2005,Basner2013,Meier2017}, but also for interstellar journeys 
to nearby exoplanets \cite{Obousy2011,Edwards2013,Gilster2013,Marin2019}. 

Sending humans to exo-worlds situated at several (tens of) parsecs away from the Earth is a rather complex project, 
since it requires to travel in deep space for hundreds of years with subluminal propulsion systems. The technological 
complexity of such giant spaceship has been highlighted by many authors \cite{Obousy2011,Hein2012,Nygren2015}, but it is 
universally recognized that the human aspect of the mission is even more complex. Multi-generational spaceship are probably 
the best option we have since population resources consumption over time can be modelled with great precision \cite{Jones2009}. 
Those generation ships rely on the principle that the population aboard will live, procreate, teach the new generation, and die, 
allowing the offspring to continue the journey. Inside the vessel, a closed ecological system would artificially reproduce 
Earth-like conditions, landscapes and flora. The population of a space ark should not be considered as a unique crew with a
distinct goal, but rather as families and communities living out normal, small-town lives in the world ship \cite{Smith2019}. 
However, in order to continue the space travel, their basic physiological needs must be fulfilled.

Among the several solutions that have been postulated to solve the issue of the inevitable depletion of food, water 
and breathable gases supplies, the most convincing one is the production of resources within the spaceship. This can be 
achieved by either recycling wastes, by growing food and plants in biospheres, or by using chemical reactions to transform 
non-breathable gases to oxygen. But how much food, water and oxygen is needed to ensure that the crew will have enough 
resources to live a prosperous life? Precise quantitative estimations have been achieved in the case of short duration
spaceflights with a limited crew (see, e.g., \cite{Mason1980}) but never in the case of long-duration interstellar travels with 
a dynamic population, which is the goal of this paper. We addressed the question of food production and requirements in a 
previous publication \cite{Marin2019}. We now turn our attention and our numerical simulation tool HERITAGE towards the issue 
of water and air consumption aboard an interstellar spaceship whose crew consists of a multi-generational population. In 
the following, we will present the latest upgrades of our agent-based code that were necessary to properly compute the annual
basal oxygen consumption and the estimated water requirements of the population. Then, we will review the various methods 
to refill the spaceship with pure water and air before concluding on the importance of planning for a long-duration space voyage.

\section{Upgrading HERITAGE}
\label{Improvements}

\subsection{Overview of the agent based Monte Carlo code}
\label{Improvements:reminder}

The numerical code HERITAGE is a computer program that was created to follow the evolution of a multi-generational 
population within a closed environment with limited resources and neither immigration nor emigration. HERITAGE has been 
applied to interstellar space travels but can very well be applied to local (Earth) situations such as an isolated 
tribe in the jungle, an island or a Mars colony experiment such as the Mars-500 project \cite{Basner2013} or the 
Hawaii - Space Exploration Analog and Simulation (HI-SEAS) \cite{Musilova2019}.The uniqueness of the code comes from 
the fact that it is an agent based tool. Each crew member is fully simulated using a specific blueprint (a C++ language class) 
that includes the most important biologic and anthropometric data that are needed to characterize a real human: age, gender, 
weight, height, genome, fertility, etc. The biological factors are time-dependent and follow biological and physical
laws so that the population can grow old, reproduce and die in a biologically realistic way. This allows us to model a real 
population with mixed generations and heterogeneous characteristics rather than populations with clearly separated 
generations, as what is usually done in population genetics studies \cite{Hamilton2009,Shukla2009}. HERITAGE is 
also a Monte Carlo code, which means that each event happening in the spaceship (reproduction, accidents, genetic processes, 
and so on) are the result of random draws that follow mathematical laws. This means that the code can test 
all possible outcomes of an event by performing successive draws. The code must be lopped several times (at least one 
hundred times, see \cite{Marin2019}) to have statistically significant results and determine the most probable outcomes.
The results of the simulations are then averaged over hundreds of loops that, depending on the number of crew members, 
can take hours to days to complete.

The code has been extensively described in the previous papers of the HERITAGE project \cite{Marin2017,Marin2018,Marin2019}.
In the following, we will only review the new features that are necessary to determine air and water consumption 
aboard a closed environment.

\subsection{Equations for air consumption}
\label{Improvements:air}

\begin{figure}
\centering
\includegraphics[trim = 0mm 0mm 0mm 0mm, clip, width=8.2cm]{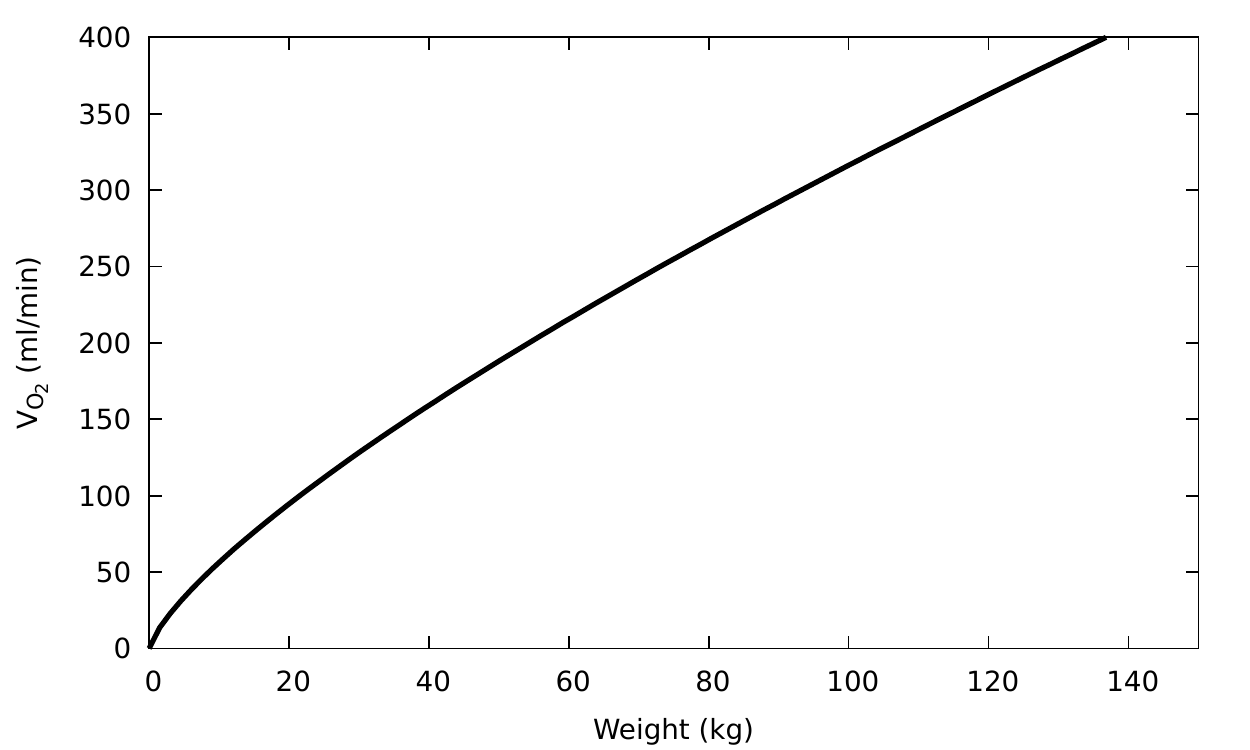}
  \caption{Plot of the Brody's equation.}
  \label{Fig:Brody}
\end{figure}

The daily quantity of air needed by a human under minimal psychological and physiologic stress, and at an ambient 
temperature comprised between 20 and 26.6$^\circ$C, has been derived thanks to indirect calorimetry during clinical 
experiments \cite{Kleiber1947,Guyton1986}. Surprisingly, this quantity depends only marginally on the sex and height 
of the test subject but more strongly on its weight. Indeed, there is a strong correlation between the body size and 
metabolic rate of mammals \cite{Kleiber1932}. Based on clinical observations, an equation was derived to express 
the basal oxygen consumption ($V_{\rm O_2}$) as a function of the subject weight $w$ \cite{Brody1945,Kleiber1947}: 
\begin{equation}
V_{\rm O_2} = 10w^{\frac{3}{4}} 
\label{Brody}
\end{equation}
in milliliter per minute. This equation is known as Brody's equation, whose representation can be found in Fig.~\ref{Fig:Brody}.
It is a continuous function that can be used as a good proxy for human oxygen needs in stress-free situations. By volume, 
dry air ($>$ 10\% humidity) contains 78.09\% nitrogen, 20.95\% oxygen, 0.93\% argon, 0.04\% carbon dioxide, and small 
amounts of other gases \cite{Cox2000}. Air also contains a variable amount of water vapor, on average around 1\% at sea
level, and 0.4\% over the entire atmosphere. From Brody's equation, it is thus possible to estimate the basal oxygen 
consumption of a whole population (if the weight of each individual is known) and then derive the related volume of 
accompanying gases.

\subsection{Equations for water consumption}
\label{Improvements:water}

Water is the largest constituent of the human body and can be viewed as the most essential nutrient. It contributes to 
the global health of humans and also accounts for a large fraction of their body weight. Ingesting insufficient daily water
doses leads to dehydration and volume depletion \cite{Thomas2008}, while too much water leads to over-hydration and 
hyponatremia \cite{Grandjean2003}. In order to avoid those life-threatening conditions, an appropriate amount of water 
must be consumed per day. It is recognized that a minimum of one liter per day is necessary to survive \cite{NRC2005}, but
if we want to move away from this scarcity assumption, it becomes less trivial to properly assess the amount of water
that is required to maintain a healthy body. There are at least five equations used by clinicians to determine water 
requirements. Studies have pointed out that those five Estimated Water Requirements (EWR) equations are strongly 
correlated but do not necessarily agree with each other \cite{Tannenbaum2011}. To determine the best equation, the 
calculated EWR results were compared to measurements from the total water intake (from food and beverages) from a 
representative sample of the United State population. The strongest agreement between the total water intake and the 
estimated water requirements is found in the case of the equation issued by the NRC (National Research Council) 
\cite{Tannenbaum2011}:
\begin{equation}
		\begin{split}
			{\rm EWR}_{\female}~=~&354 - (6.91 \times A) + (9.36 \times W \times PAL) \\
			&- (726 \times H \times PAL)
		\end{split}
\label{Harris-Benedict_wom}
\end{equation} 
\begin{equation}
		\begin{split}
			{\rm EWR}_{\male}~=~&662 - (9.53 \times A) + (15.91 \times W \times PAL) \\
			& - (540 \times H \times PAL)
		\end{split}
\label{Harris-Benedict_man}
\end{equation} 
with $W$ the weight in kilograms, $H$ the height in meters, $A$ the age in years, and the EWR in milliliter per day.
The acronym PAL stands for Physical Activity Level and is used to express a person's daily physical activity as a
quantitative value. Those equations are based on the basal metabolic rate of individuals, such as already computed by
HERITAGE (see \cite{Marin2019}). They have the advantage of accounting for the body mass, body height, age, gender and 
physical activity level of the subject. All those parameters are already included in HERITAGE from the previous 
upgrade. We thus simply included those equations into our code to estimate the daily water requirements. 

\begin{figure}
\centering
\includegraphics[trim = 0mm 0mm 0mm 2mm, clip, width=8.2cm]{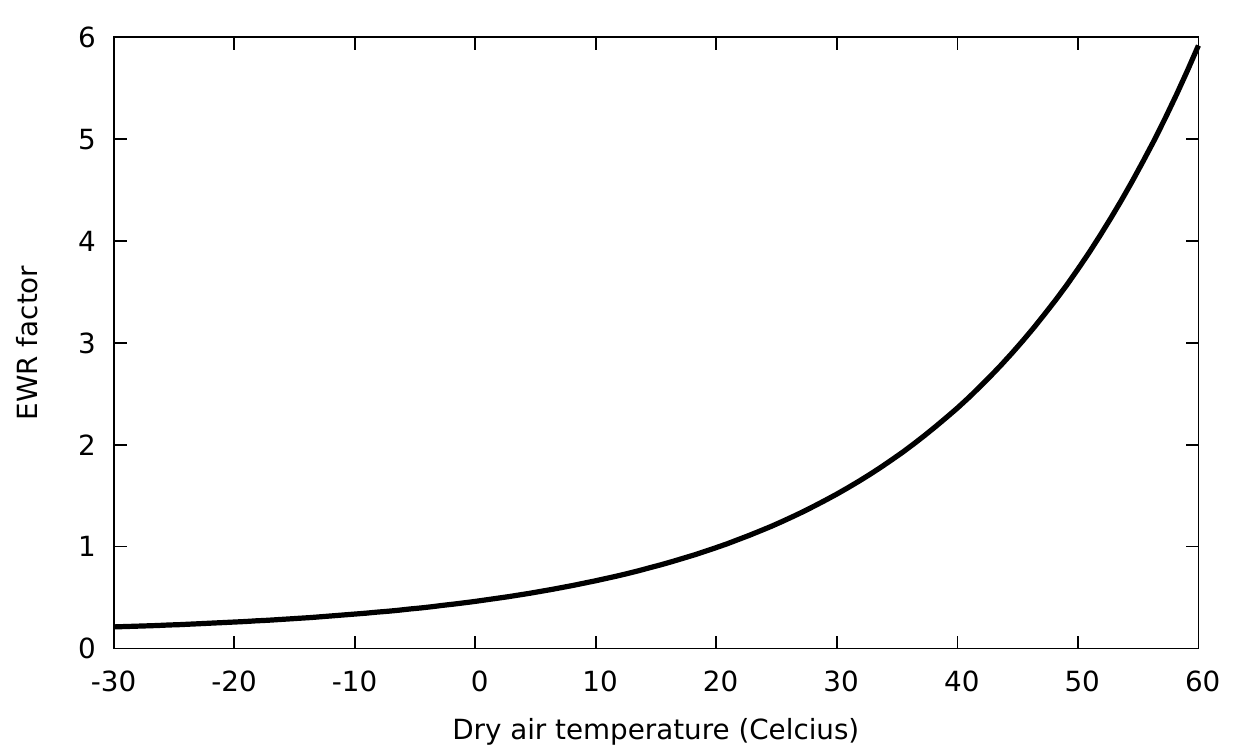}
  \caption{Correction factor applied to the EWR as a function of dry 
	  air temperature aboard the spaceship. The model is normalized
	  to unity at a nominal temperature slightly below 20$^\circ$C 
	  and detailed in the text.}
  \label{Fig:Temperature}
\end{figure}

It must be noted that the estimated water requirements are very temperature-sensitive: EWR increases with ambient 
temperature as a result of sweating. The United States Army Research Institute of Environmental Medicine has developed 
an empirical model that includes an equation to predict sweating rate during work \cite{Moran1995,Shapiro1995}.
The model is valid for dry air temperatures that range between 15 and 40$^\circ$C and accounts for four different 
physical activity levels. The model has an exponential dependency on temperature with proportional rate growths 
(see \cite{NRC2005}). We used the model to numerically determine a temperature-dependent correction factor 
$f_{\rm EWR}$ to apply to the EWR value obtained with Eqs.~\ref{Harris-Benedict_wom} and \ref{Harris-Benedict_man}:
\begin{equation}
f_{\rm EWR} = a - \frac{b}{c} \times (1-exp(c \times T))
\label{EWR_correction}
\end{equation}
with $T$ the dry air temperature in Celsius degrees, $a$ = 0.462, $b$ = 0.016 and $c$ = 0.048. This
correction factor must be multiplied to the EWR to account for the temperature-dependent daily dose of water
needed by each crew member. A visual representation of $f_{\rm EWR}$ is shown in Fig.~\ref{Fig:Temperature}.

The correction factor has been normalized to an ambient dry air temperature slightly below 20$^\circ$C and a numerical 
fit allowed us to extend the temperature range from negative values up to more than 60$^\circ$C. Nevertheless, 
the HERITAGE code will emit a warning message whenever the temperature within the spaceship is below or above
the nominal temperature range fixed by \cite{Moran1995,Shapiro1995}, so that the user will know that the results
are subject to larger uncertainties. The annual dry air temperature aboard the spaceship is now input data (similarly
to the year-by-year catastrophe risks and equivalent doses of cosmic ray radiation) that can be fixed by the user. 
Note that for climate control, a typical ambient temperature range is anywhere between 15 and 25$^\circ$C. The human
body, dressed appropriately, can survive for a year under extreme conditions ranging from -60 to +50$^\circ$C
in dry air conditions and with regular access to water at cooler temperatures in the latter case \cite{Kellogg2019}. 
Above and below those limits, the crew members are killed by the program. Finally, effects of extreme temperatures 
on the human biological functions, such as fertility or blood flow \cite{Seltenrich2015,Coccarelli2017}, are rather
complex to model and are not included in HERITAGE at this point.

\section{Simulations}
\label{Simulations}

\begin{table*}[!t]
  \centering
    \begin{tabular}{lrl}
	\textbf{Parameters} & \textbf{Values} & \textbf{Units} \\
        \hline
        Number of space voyages to simulate & 100 & -- \\
	Duration of the interstellar travel & 600 & years \\
	Colony ship capacity & 1200 & humans \\
	Overpopulation threshold & 0.9 & fraction \\	
	Inclusion of Adaptive Social Engineering Principles (0 = no, 1 = yes) & 1 & -- \\	
	Genetically realistic initial population (0 = no, 1 = yes) & 1 & -- \\
	Number of initial women & 250 & humans \\
	Number of initial men & 250 & humans \\
	Age of the initial women & 30 $\pm$ 5 & years \\
	Age of the initial men & 30 $\pm$ 5 & years \\
	Number of children per woman & 2 $\pm$ 0.5 & humans \\
	Twinning rate & 0.015 & fraction \\
	Life expectancy for women & 85 $\pm$ 5 & years \\
	Life expectancy for men & 79 $\pm$ 5 & years \\
	Mean age of menopause & 45 & years \\
	Start of permitted procreation & 18 & years \\
	End of permitted procreation & 40 & years \\
	Initial consanguinity & 0 & fraction \\
	Allowed consanguinity & 0 & fraction \\
	Life reduction due to consanguinity & 0.5 & fraction \\	
	Chaotic element of any human expedition & 0.001 & fraction \\
        \hline
    \end{tabular}
    \caption{Input parameters of the simulation. The $\mu$ $\pm$ $\sigma$ values shown for
	    certain parameters indicate that the code needs a mean ($\mu$) and a 
	    standard deviation value ($\sigma$) to sample a number from of a normal 
	    (Gaussian) distribution.}
    \label{Tab:Parameters}
\end{table*}

\begin{figure*}[!t]
  \begin{subfigure}[t]{.48\textwidth}
    \centering
    \includegraphics[width=\linewidth]{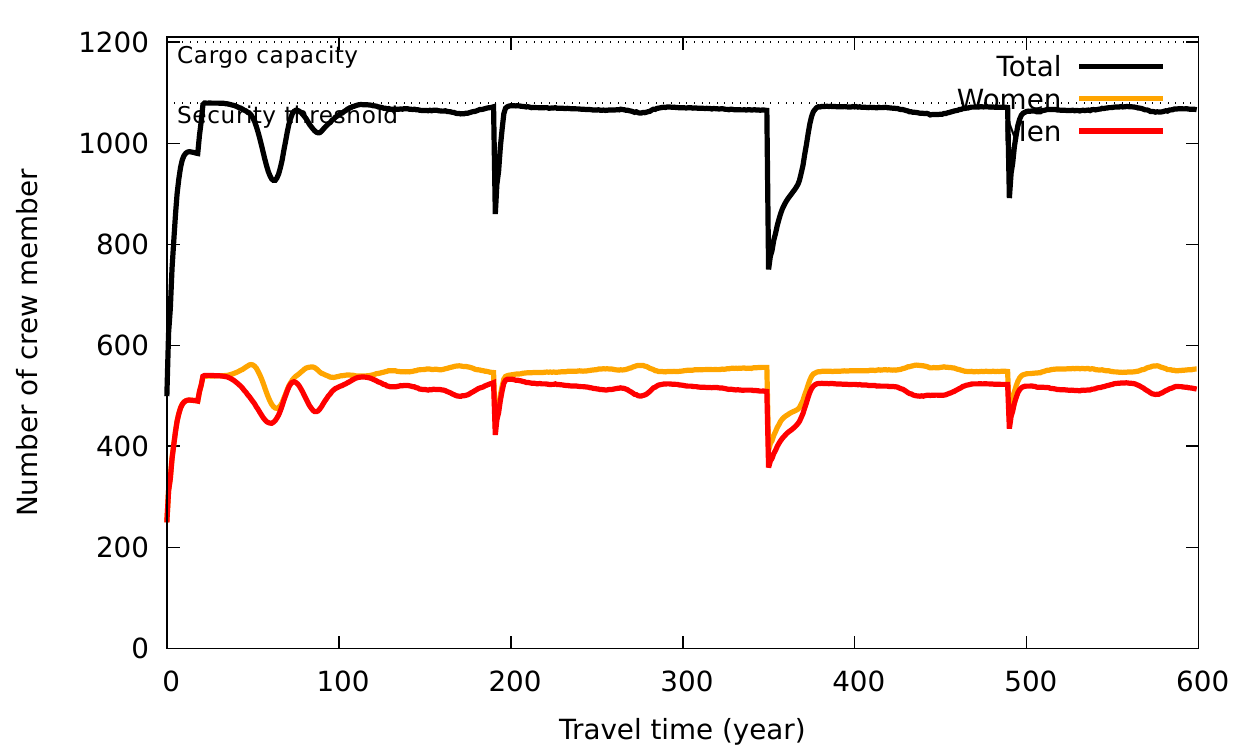}
    \caption{Crew evolution in terms of population number (black: total,
    orange: women, red: men).}
  \end{subfigure}
  \hfill
  \begin{subfigure}[t]{.48\textwidth}
    \centering
    \includegraphics[width=\linewidth]{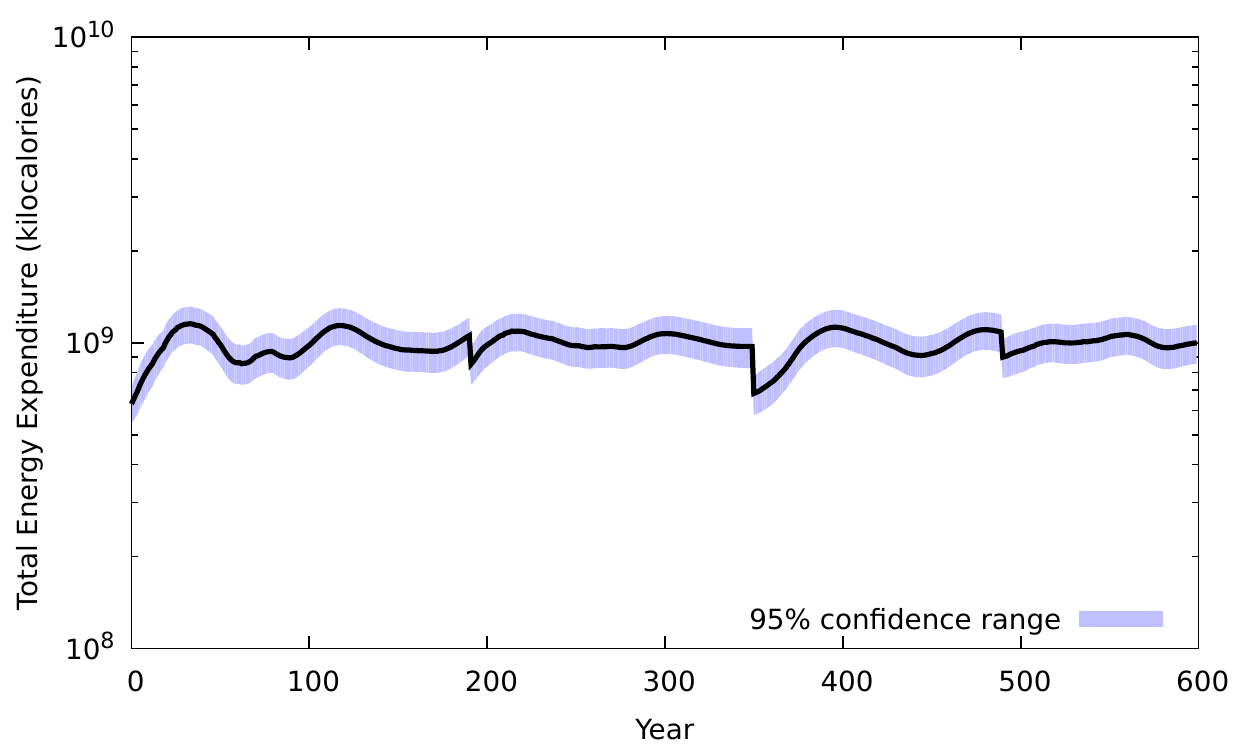}
    \caption{Total energy expenditure (in kilo-calories) per year.}
  \end{subfigure}

  \medskip

  \begin{subfigure}[t]{.48\textwidth}
    \centering
    \includegraphics[width=\linewidth]{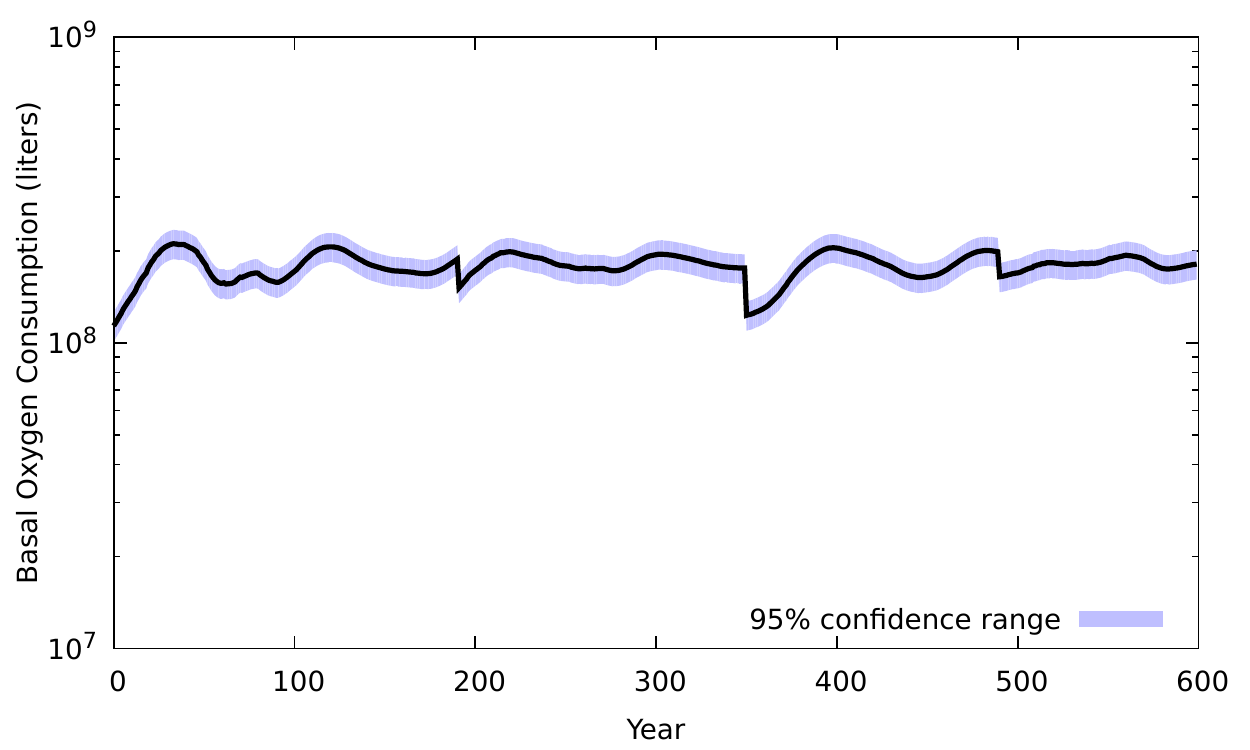}
    \caption{Basal oxygen consumption (in liters) per year.}
  \end{subfigure}
  \hfill
  \begin{subfigure}[t]{.48\textwidth}
    \centering
    \includegraphics[width=\linewidth]{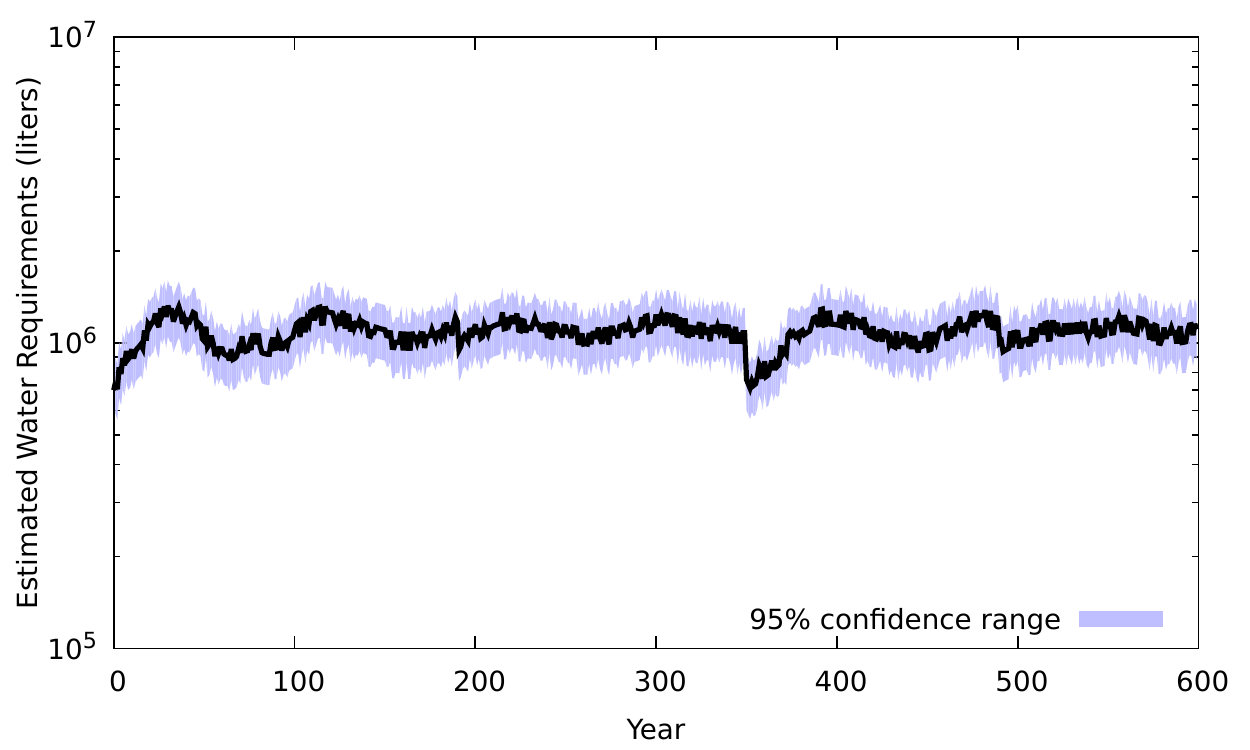}
    \caption{Estimated water requirements (in liters) per year.}
  \end{subfigure}
  
  \caption{HERITAGE results for a 600 years-long interstellar travel under 
	  the conditions described in the text. The supplemental noise seen 
	  in the estimated water requirements is due to the random fluctuations
	  of the dry air temperature aboard the spaceship.}
  \label{Fig:Simulation}
\end{figure*}

Now that the code has been upgraded to account for air (oxygen) and water consumption, we can run a complete 
simulation to obtain the annual requirements in terms of food, water and air. To do so, we set up HERITAGE 
according to the parameters listed in Tab.~\ref{Tab:Parameters}. We simulate a relatively short space travel
(600 year long) with a worldship of modest size: a maximum of 1200 crew members can live inside the vessel but
we impose a security threshold of 90\% of that value to prevent overpopulation and resource depletion. We include
adaptive social engineering principles that is a self-regulation of the population whenever the overpopulation
threshold is reached. No inbreeding is tolerated on board (up to first cousins once removed or half-first cousins) 
and the procreation window is wide: from 18 to 40 years old. The crew is expected to be rather inactive when young 
($<$ 18 years old), moderately active in their age 18 -- 24, vigorously active between 25 and 50, moderately 
active up to 70 years old and sedentary for the rest of their life. We postulate that the radiation shield of the
interstellar ark is very efficient so that the annual equivalent dose of cosmic ray radiation never exceeds
2 milli-Sieverts in deep space (i.e., beyond the protective limits of the Sun and of the star around which is 
orbiting the targeted exoplanet). This prevents any serious risks of neo-mutations, cancers and genetic 
malformations. The initial crew is gender-balanced and older than in our previous simulations: 30 years old 
on average with a standard deviation of 5. This means that a few crew members are older than 40 years old and 
some are younger than 20 at the beginning of the mission. To move away from the scarcity paradigm (i.e., imposing
the minimal viable population of initial settlers at year 0, which is of the order of 100, see \cite{Marin2018}),
we choose a departing population of 500 humans. We will discuss the socio-demographic importance of the initial 
population age and number in another publication. We set up a chaotic factor that can randomly kill any 
crew member for unexpected reasons (deadly accident, premature death, serious illness, etc.) to 0.1\% per
year. Finally, the dry air temperature aboard is stochastically oscillating between 18 and 21$^\circ$C and we 
set up three incidents along the course of the multi-generational ark at years 191 (20\% causalities), 350 
(30\% causalities) and 490 (17\% causalities). The casualty percentage rates are chosen to mimic large-scale
catastrophes and diseases, such as the 14th century Black Plague ($\sim$ 30\% of the European population died, 
\cite{Hollingsworth1969}) or the 1918 -- 1919 Spanish flu pandemics ($\sim$ 22\% of the Samoa population died,
\cite{Kohn2007}).

The results of the simulations are shown in Fig.~\ref{Fig:Simulation}. The top-left panel shows the evolution of 
the population within the spacecraft. From 500 at the beginning of the interstellar travel, the crew quickly 
increases up to the security threshold. At this point, the population number is subject to strong variations as 
the crew is dominated by demographic echelons clustered in discrete age groups. With time, this clustering gradually
disappears and a stable population level with people of many different ages develops. The three catastrophes have little 
impact onto the resilience of the crew and the population quickly returns to the stable population level after the lethal 
events. The rate of mission success is precisely 100\% in this case. The top-right panel of Fig.~\ref{Fig:Simulation} shows
the total energy expenditure in kilo-calories per year. On average, there is a need for $\sim$ 10$^9$ kcal per year to 
maintain ideal body weight. Following our investigations presented in \cite{Marin2019}, this total energy expenditure 
drives a surface for geoponic agriculture of about 2 square kilometers of farmland for a balanced diet (fruits, vegetables, 
meat and fish, dairy and starch). The use of hydroponic and/or aeroponic agricultures could help to reduce this surface 
by a factor two. The bottom-left figure presents the basal oxygen consumption in liters per year. We see that 
about 1.8 $\times$ 10$^8$ liters of oxygen are annually required to keep the crew alive. At the spaceship ambient dry
air temperature, considering a pressure of one atmosphere, this represents a volume\footnote{1 liter of oxygen gas 
represents a volume of 10$^{-3}$~m$^3$ and a mass of 1.309~g at 21$^\circ$C and 1~atm.} of 1.8 $\times$ 10$^5$ 
cubic meters (2.4 $\times$ 10$^5$ kilograms). If we account for nitrogen among the breathable gases, at a fraction of 
78.09\% of the air composition, this means that an additional amount of 6.7 $\times$ 10$^8$ liters of nitrogen is needed 
inside the vessel, which corresponds\footnote{1 liter of nitrogen gas represents a volume of 10$^{-3}$m$^3$ and a mass 
of 1.155~g at 21$^\circ$C and 1~atm.} to $\sim$ 6.7 $\times$ 10$^5$ cubic meters (7.7 $\times$ 10$^5$ kilograms). 
In total, the spaceship must have a volume of about 10$^6$ cubic meters just to store the required breathable gases 
(nitrogen and oxygen) at 21$^\circ$C and at a pressure of one atmosphere for a full year. Considering the simplest 
geometry for such spaceship (a cylinder), this represents a structural length longer than 350 meters for a fixed radius 
of 30 meters. Storing a fraction of the gases in pressurized tanks would help to drastically decrease such architectural
constraints. The breathable gases could then be released inside the vessel according the monitored needs of the crew. 
In addition, it is not necessary to keep enough gases for a full year. Day-to-day oxygen and nitrogen production aboard
must participate actively in the renewal of the breathable atmosphere (see Sect.~\ref{Productions:Air}). Finally, the 
bottom-right panel of Fig.~\ref{Fig:Simulation} presents the estimated water requirements in liters per year. There is
an annual need of 1.1 $\times$ 10$^6$ liters of water to hydrate the population. Of course, this amount of water does 
not include the volume required by the biosphere (plants, animals, insects, bacteria ...) inside the ark to survive. 
It gives us a lower limit solely based on human needs. This quantity represents a volume of 1100 cubic meters of water 
(1.1 $\times$ 10$^6$ kilograms), or a cylinder of length 35.4 meters and radius 3 meters. This is a much smaller volume
than what is required by breathable gases requirements; water could be stored in containers that are fixed on the 
vessel's flanks, minimizing the loss of space inside the spacecraft. Due to the coldness of the interstellar atomic and 
molecular media ($\le$ -173.15$^\circ$C, \cite{Lequeux2004}), water stored outside the spaceship would freeze (no need for
active cooling in deep space) and could be used as an ice shield against interstellar debris. Pathogens would not proliferate 
at those temperatures and sterilizing water is a known and easy process. In principle, collecting the required amounts of water
from Solar System sources could be achieved well before the mission even begins but, from cost-wise and security reasons, 
it would be preferable to obtain new and fresh water during the journey. The question of how to renew the water supply will 
be discussed in Sect.~\ref{Productions:Water}.

\section{Producing air and water resources in space}
\label{Productions}
Our HERITAGE simulations have proven that tremendous annual amounts of oxygen and water are required to keep the crew alive during 
centuries-long deep space travels. It is not a viable option to store all the necessary resources at the beginning of the travel 
since the quality of liquid water would likely deteriorate with time. Microbial cells (pathogenic bacteria such as, e.g., legionella) 
would ultimately grow and induce pathogenic properties. Excessive growth of bacteria in drinking water leads to hygienic, aesthetic 
and operational problems \cite{Zhang2019}. In addition, the required volumes to store millions of water litres and billions of cubic 
meters of breathable gases would induce disproportionate costs that could affect the overall feasibility of ark-like projects. For
those reasons, \textit{in situ} recycling and production of fresh air, water and food are necessary \cite{Marin2019}.

\subsection{How to recycle and produce breathable gases?}
\label{Productions:Air}
The most convincing and accurate examples that we have concerning the production of breathable gases in space are the Mir and ISS
space stations. Inside those orbiting habitats, large scale plant photosynthesis is not feasible to produce enough oxygen so they rely on 
several methods to generate breathable gases: pressurized oxygen tanks, oxygen generators and solid fuel oxygen generators. These systems
are not perfectly efficient and losses are compensated by deliveries from Earth. Of course, the delivery of oxygen to an interstellar 
spaceship is not realistic, but nothing prevents the crew to store pressurized tanks in case of a disaster. To provide a continuous flow 
of oxygen, generators such as the Russian-made Elektron-VM \cite{Duchesne2010,Proshkin2014} and the United States Environmental Control 
and Life Support System (ECLSS, \cite{Shkedi2004,Bagdigian2005}) are used. The oxygen generators option relies on water electrolysis 
that splits H$_2$0 molecules into hydrogen gas and oxygen gas such as:
\begin{equation}
2 H^+ + 2e^- \rightarrow H_2 
\end{equation}
at the cathode and:
\begin{equation}
2 H_2O \rightarrow O_2 + 4 e^- + 4 H^+
\end{equation}
at the anode when the half reactions are balanced with acid. In the case of a balance with a base, the reactions become:
\begin{equation}
2 H_2O + 2e^- \rightarrow H_2 + 2 OH^-
\end{equation}
at the cathode and:
\begin{equation}
2 OH^- \rightarrow \frac{1}{2} O_2 + H_2O + 2 e^-
\end{equation}
at the anode. Combining either half reaction pair yields the same overall decomposition of water into oxygen and hydrogen:
\begin{equation}
2 H_2O \rightarrow 2H_2 + O_2
\end{equation}

The electricity is generated by the station's solar panels and supplied to the oxygen generators through the station's power grid 
\cite{Hale2011}. Water is essentially provided from Earth by space shuttles, which make oxygen generators such as Elektron-VM
and ECLSS less reliable in case of water shortage. The last method to produce breathable gases uses chemical reactions. Solid fuel oxygen 
generators consist of canisters that contain a mixture of powdered sodium chlorate (NaClO$_3$) and iron (Fe) powder. The ignition of the 
iron powder provides the heat energy required to break down the sodium chlorate into sodium chloride and oxygen gas (plus residual 
iron oxide) such as: 
\begin{equation}
NaClO_3 + Fe \rightarrow 3 O_2 + NaCl + FeO.
\end{equation}

One kilogram of mixture is enough to provide oxygen for 6.5 man-hours \cite{Chan2002,Podgorodetskii2015,Chen2017,Liu2019}. Chemical 
oxygen generators are the most used methods to supply oxygen in confined spaces, especially because of their long shelf life
and reliability. Nevertheless, all the three technologies require resources that are not effortlessly accessible in deep space. 
Recycling breathable gases is necessary but not easily achievable, even in space shuttles and space stations. Experiments involving
organic compounds \cite{Bockstahler2013,Putz2016} or microalgae \cite{Binot1994,Gros2003,Tikhomirov2003} to recycle carbon dioxide 
are under study but they are not expected to be 100\% efficient. The only way to provide enough breathable gases and recycle gas 
wastes is to mimic the Earth system: growing plants in space. This is a challenging task but the recent success of the Chang'e~4 
lunar lander in January 2019 is paving the way towards more and more complex biosphere experiments. The Chinese mission carried a 
3~kg sealed biosphere-like box with seeds and insect eggs to test whether plants and insects could hatch and grow together in 
synergy. It successfully achieved the sprout of cottonseed (and, maybe, of rapeseed and potato seeds, but this has not been confirmed
yet) within nine days, before a failure of the temperature system \cite{Jones2019}. Larger biosphere experiments, such as Biosphere~II,
are trying to recreate a viable ecosystem inside a huge closed dome that could very well be representative of a spacecraft 
\cite{Petersen1992,Nelson1993,Nelson1995,Alling2005}. This option is the only realistic one to consistently provide enough breathable
gases to multi-generational crews, recycle carbon dioxide and other by-products of life aboard a closed system, and create a pleasant
living and working environment for the crew. However, this potential solution must be explored carefully since the carbon dioxide
content of the air is low; hence it might require a significant extra mass of carbon. In addition, it is extremely difficult 
to calculate how the efficiency works with such complicated networks, since some carbon dioxide will be taken up by the plants for 
photosynthesis. Large scale experiments, such as Biosphere~II, are therefore fundamental to the preparation of interstellar arks.

\subsection{Where to find water in deep space?}
\label{Productions:Water}
The problem of renewing water in space is even more complex than that of creating oxygen in a closed environment. The Solar system
is full of small icy bodies (comets and asteroids) that could be collected prior to the mission or even during the spaceflight to 
restore the water tanks level after purification of the melted ice. Harvesting water in the Solar system before the launch of the 
space ark might be a good option since it can be done during the construction time of the spaceship. However, as we already stated it, 
it is dangerous to travel with all the water stored for the mission. In case of a catastrophe that would destroy part of the 
reserves, this would mean the end of the mission. It is then necessary to collect additional water on the course to the exoplanet.
But, is this really an option? The closest asteroid belt is the Main Belt. It is estimated to contain millions of objects 
and is located between the orbits of Mars and Jupiter ($\sim$ 2 -- 3 astronomical units, \textit{au}, from the Sun). Further away are 
Trojan asteroids, rocky/icy bodies that are a separate group of asteroids lying outside the main asteroid belt and sharing an orbit 
with Jupiter ($\sim$ 5 -- 6~au). Several comets and ice-coated irregular satellites of the giant planets are situated shortly after 
the Trojan asteroids, up to the Neptune Trojans at $\sim$ 30~au. At larger distances from the Sun, Kuiper belt objects fill the space 
up to $\sim$ 55~au. Most of the aforementioned asteroids are situated along (or close) to the planet's orbital plane \cite{Vernazza2016}. 
This means that any interstellar travel whose direction does not lie close to the Solar system ecliptic plane is less likely to encounter
asteroid that could be harvested to restore water levels (at least up to 55~au). In addition to the inherent risks of boarding a fast 
moving\footnote{The average orbital speed of a main-belt asteroid is 17.9~km.s$^{-1}$, the orbital speed of Ceres \cite{Park2016}. 
Ceres has a pretty typical orbit and makes up a third of the Belt by mass \cite{Platz2016,Prettyman2017}.} space rock, the chances to 
encounter enough asteroids along the spaceship's trajectory within the Solar system is rather low if the ship's path deviates from the 
ecliptic. The only true potential reservoir of frozen water that lies ``close'' to the Sun is the Oort cloud. The Oort cloud is a large 
hypothetical spherical reservoir of comets that surrounds the Solar system at distances between 20\,000 and $>$ 100\,000~au \cite{Correa2019}. 
The outer envelope could be the source of most long-lived comets and be an ideal target for water replenishment if the risk is worth it.
Finally, in the deep interstellar space, the number density of comets and asteroids that could be used to extract water is so low 
(1.4 $\times$ 10$^{-4}$~au$^{-3}$ at a 90\% confidence limit, \cite{Engelhardt2017}) that such an option could be safely discarded.

We established that extracting water from comets and asteroids during the spaceship course is not a viable option. Is recycling water 
from humans, plants and industrial activities within the spaceship feasible? Aboard the ISS there are two water recovery systems, one
that processes water vapor from the atmosphere (the water is then fed to electrolysis oxygen generators) and one that processes water
vapor collected from the atmosphere and urine into water that is intended for drinking. However, they are not 100\% efficient since 
some water is lost due to small amounts of unusable brine produced by the recycling systems, water consumption by the oxygen generators, 
airlocks leaks that take humidity with them, and a few more. Even with a 95\% efficient system, such as the NASA's Vapor Compression 
Distillation experiment \cite{Zdankiewicz1985}, the amount of water that is lost per year is non-negligible and requires filling the
reserves with fresh water. This demonstration has the corollary that some water must be created within the ship in addition to the 
water recovered from the recycling systems. To do so, we can rely on chemical reactions. In particular, the Sabatier process could be 
used to produce methane and water. To do so, the methodology relies on the reaction of hydrogen with carbon dioxide at elevated temperatures 
(optimally 300 -- 400~$^\circ$C) and pressures in the presence of a nickel catalyst \cite{Ronsch2016,Luo2017,Meier2017}. The chemical 
exothermic process is such as:
\begin{equation}
CO_2 + 4 H_2 \rightarrow CH_4 + 2H_2O.
\end{equation}
A very interesting by-product of the Sabatier reaction is the production of methane that could be very well used as a propellant
for the spaceship. In addition, this process would recycle the human production of carbon dioxide to produce water, that in turn 
could be used to water the plants and create a complete loop. Along the same line of thoughts, the Bosch reaction is under study 
to complement the Sabatier reaction to maximize water production aboard the ISS or during future Martian colonies \cite{Meier2017}. 
The overall reaction is:
\begin{equation}
CO_2 + 2 H_2 \rightarrow C + 2H_2O.
\end{equation}
This reaction requires the introduction of iron as a catalyst and requires a temperature range of 530 -- 730~$^\circ$C \cite{Messerschmid2013}, 
which is substantially higher than the Sabatier process. Yet, similarly to the first chemical method, the Bosch reaction would 
also allow for a completely closed hydrogen and oxygen cycle within the biosphere. The by-product of this reaction is the production
of carbon that could be used for manufacturing, but the production of elemental carbon tends to foul the catalyst's surface, which is 
detrimental to the reaction's efficiency.

\subsection{Life support systems, virtual habitat models and HERITAGE}
\label{Productions:Models}
The ECLSS aboard the ISS is part of the more general class of life support system (LSS). The goal of any LSS is to create and manage a viable 
environment with sufficient breathable gases, water and food, but also to maintain human-adapted temperature and pressure conditions. 
Another task of a LSS is to manage waste and recycle unnecessary material. The ECLSS is crucial for the survivability of the ISS 
crew and it is easy to foresee that a similar (yet at larger scale) situation would apply in the case of generation ships. LSS 
are well adapted for astronaut crews that are, on average, anthropometrically similar. In the case of the ECLSS, the human baseline 
values and assumptions are such: a typical crew member has a mass ranging from a 95$^{\rm th}$ percentile American male, with a total
body mass of 99~kg, to a 5$^{\rm th}$ percentile Japanese female, with a total mass of 53~kg \cite{Anderson2015}. Despite the use of 
similar equations for the basal metabolic rate and water consumption than in this paper, the ECLSS assumptions about the crew body
mass and metabolic rate are only valid for humans older than 19 years. The water/air consumption and wastes of the same standard crew member 
with a fixed respiratory quotient at a fixed ambient temperature are averaged, and the physical activity levels are planned over a 
standard workweek \cite{Langston2012}. No children nor elder crew members are accounted for in the human model of the ECLSS. Those 
strong assumptions are perfectly valid for the ISS case, where only trained astronauts are likely to go, but the assumptions must
be corrected in the case of an interstellar ark. Our HERITAGE code is certainly not as efficient as the ECLSS model in the case 
of a daily estimation of the human needs. Most of the biological functions included in the code are based on medical data that are 
year-dependent, not day-dependent. For this reason, HERITAGE is not likely to replace a LSS. However, our code has the dual benefit 
of being adapted to the yearly consumption of any type of population and it is a dynamical code that reacts to changes in temperature, 
cosmic ray radiation doses, population size variations, etc.

HERITAGE shares more comparative points with virtual habitat models. In particular, the V-HAB model developed by the Exploration Group 
in Munich \cite{Czupalla2007,Plotner2013,Czupalla2015} includes a dynamic representation of the crew to be integrated in any LSS. 
The human model includes a much more sophisticated representation of the human body (lungs, heart, kidneys, fluids ...) and very precise 
estimations of the water, food and breathable gases intake can be achieved. The V-HAB model works very well when compared to ISS data
and can be used to prepare a Mars colony with reliable numbers. Nevertheless, the human model in V-HAB remains static, in the sense that 
only standardized, active, male adults can be modelled \cite{Czupalla2015}. HERITAGE accounts for gender, age, size, weight and genetic 
diversity. This makes HERITAGE a valuable and complementary tool to LSS and virtual habitat models to obtain precise (at a year timescale) 
quantitative values of the human needs in terms of food, water and breathable gases, accounting for a dynamical population that is 
truly representative of an off-world colony.

\section{Conclusions}
\label{Conclusions}
We have developed the HERITAGE code to include water and oxygen consumption aboard multi-generational interstellar spaceship. 
The code now estimates the annual requirements in terms of water and oxygen volumes that a heterogeneous population would need to 
live comfortably, accounting for various physical activity levels, body shapes and ambient dry air temperatures. The results 
of a simulation where about 1\,100 crew members are numerically investigated shows that about 1.8 $\times$ 10$^8$ liters of 
oxygen are annually required to keep the crew alive, together with $\sim$ 10$^6$ liters of water. Those results do not account 
for oxygen and water used in growing plants, but they give us an insight of how much resources are needed in the spaceship.
This, in turns, help us to determine the architectural constraints of such an enterprise.

In order to narrow down the possibilities to restore the levels of breathable gases and water, we undertook a concise review of 
the methods for recycling and producing air and water in space. The most convincing methods rely on photosynthesis aboard the 
spaceship and chemical reactions to maintain a closed loop for water and air. Humans (and animals and insects) would inhale air 
and exhale carbon dioxide. This carbon dioxide will in turn be partially recycled by the plants from the biosphere, and partially 
used to feed the Sabatier and Bosch chemical reactors to produce water. Water will be used to hydrate humans and water the plants.
The by-products of the chemical reactors could be then used as engine fuel or raw material for industries. This first order scheme 
is of course too simplistic as such, but it gives us a direction of study to really get a stable biosphere.

\section*{Acknowledgment}
The authors would like to thank the anonymous referee for her/his suggestions that help to improve the clarity of this paper.
The authors also acknowledge Dr. Rhys Taylor (Astronomical Institute of the Czech Republic) and Dr. Katharina Lutz
(Astronomical Observatory of Strasbourg) for their relevant suggestions and corrections to the original manuscript.

\bibliography{mybibfile}

\begin{thebibliography}{10}
\expandafter\ifx\csname url\endcsname\relax
  \def\url#1{\texttt{#1}}\fi
\expandafter\ifx\csname urlprefix\endcsname\relax\def\urlprefix{URL }\fi
\expandafter\ifx\csname href\endcsname\relax
  \def\href#1#2{#2} \def\path#1{#1}\fi

\bibitem{Reynaud1973}
F.~Reynaud, Le 21eme voyage du "clairon et reine", brick marseillais (1833),
  Provence historique 93-94 (1973) 445--462.

\bibitem{Reynaud1975}
F.~Reynaud, L'alimentation des marins vers 1830, Provence historique 101 (1975)
  475--486.

\bibitem{Kokini1995}
J.~Kokini, G.~Ghai, C.-T. Ho, M.~Karwe, F.~Henrikson, Food stability and shelf
  life (1995) 29.

\bibitem{Subramaniam2000}
P.~Subramaniam, D.~Kilcast, The Stability and Shelf-Life of Food, Woodhead
  Publishing Series in Food Science, Technology and Nutrition, Elsevier Science
  (2000).

\bibitem{Nelson1993}
M.~Nelson, T.~Burgess, A.~Alling, N.~Alvarez-Romo, W.~Dempster, R.~Walford,
  J.~Allen, Using a closed ecological system to study earth's biosphere:
  initial results from biosphere 2, Bioscience 43 (1993) 225--36.

\bibitem{Nelson1995}
M.~Nelson, W.~Dempster, Living in space: Results from biosphere 2's initial
  closure, an early testbed for closed ecological systems on mars, Life support
  \& biosphere science : international journal of earth space 2 (1995) 81--102.

\bibitem{Alling2005}
A.~Alling, M.~Van~Thillo, W.~Dempster, M.~Nelson, S.~Silverstone, J.~Allen,
  Lessons learned from biosphere 2 and laboratory biosphere closed systems
  experiments for the mars on earth project, Biological Sciences in Space 19
  (2005) 250--260.

\bibitem{Basner2013}
M.~Basner, D.~F. Dinges, D.~Mollicone, A.~Ecker, C.~W. Jones, E.~C. Hyder,
  A.~Di~Antonio, I.~Savelev, K.~Kan, N.~Goel, B.~V. Morukov, J.~P. Sutton, Mars
  520-d mission simulation reveals protracted crew hypokinesis and alterations
  of sleep duration and timing, Proceedings of the National Academy of Sciences
  110~(7) (2013) 2635--2640.

\bibitem{Meier2017}
A.~Meier, M.~Shah, P.~Hintze, A.~Muscatello, Mars atmospheric conversion to
  methane and water: An engineering model of the sabatier reactor with
  characterization of ru/al2o3 for long duration use on mars, 47th
  International Conference on Environmental Systems (2017).

\bibitem{Obousy2011}
R.~K. {Obousy}, A.~C. {Tziolas}, K.~F. {Long}, P.~{Galea}, A.~{Crowl}, I.~A.
  {Crawford}, R.~{Swinney}, A.~{Hein}, R.~{Osborne}, P.~{Reiss}, {Project
  Icarus: Progress Report on Technical Developments and Design Considerations},
  Journal of the British Interplanetary Society 64 (2011) 358--371.

\bibitem{Edwards2013}
M.~R. {Edwards}, {Sustainable Foods and Medicines Support Vitality, Sex and
  Longevity for a 100-Year Starship Expedition}, Journal of the British
  Interplanetary Society 66 (2013) 125--132.

\bibitem{Gilster2013}
P.~A. {Gilster}, {Slow Boat to Centauri: A Millennial Journey Exploiting
  Resources Along the Way}, Journal of the British Interplanetary Society 66
  (2013) 302--311.

\bibitem{Marin2019}
F.~{Marin}, C.~{Beluffi}, R.~{Taylor}, L.~{Grau}, {Numerical constraints on the
  size of generation ships from total energy expenditure on board, annual food
  production and space farming techniques}, Journal of the British
  Interplanetary Society 71 (2018) 382--393.

\bibitem{Hein2012}
A.~M. Hein, M.~Pak, D.~Putz, C.~Buhler, P.~Reiss, World ships - architectures
  \& feasibility revisited, Journal of the British Interplanetary Society 65
  (2012) 119--133.

\bibitem{Nygren2015}
E.~Nygren, Hypothetical Spacecraft and Interstellar Travel, Lulu.com, 2015.

\bibitem{Jones2009}
H.~Jones, Starship life support, 2009.

\bibitem{Smith2019}
C.~Smith, Principles of Space Anthropology: Establishing a Science of Human
  Space Settlement, Space and Society, Springer International Publishing, 2019.

\bibitem{Mason1980}
R.~Mason, Celss scenario analysis: Breakeven calculations (1980).

\bibitem{Musilova2019}
M.~{Musilova}, H.~{Rogers}, B.~{Foing}, N.~{Sirikan}, A.~{Weert}, S.~{Mulder},
  B.~{Pothier}, J.~{Burstein}, {EMM IMA HI-SEAS campaign February 2019}, in:
  EPSC-DPS Joint Meeting (2019).

\bibitem{Hamilton2009}
M.~Hamilton, Population Genetics, Wiley (2009).

\bibitem{Shukla2009}
A.~Shukla, Population Genetics, Encyclopaedia of genetics, Discovery Publishing
  House Pvt. Limited (2009).

\bibitem{Marin2017}
F.~{Marin}, {HERITAGE: A Monte Carlo code to evaluate the viability of
  interstellar travels using a multi-generational crew}, Journal of the British
  Interplanetary Society 70 (2017) 184--195.

\bibitem{Marin2018}
F.~{Marin}, C.~{Beluffi}, {Computing the Minimal Crew for a multi-generational
  space journey towards Proxima b}, Journal of the British Interplanetary
  Society 71 (2018) 45--52.

\bibitem{Kleiber1947}
M.~Kleiber, Body size and metabolic rate, Physiological Reviews 27~(4) (1947)
  511--541.

\bibitem{Guyton1986}
A.~Guyton, Textbook of medical physiology, Saunders (1986).

\bibitem{Kleiber1932}
M.~Kleiber, et~al., Body size and metabolism, Hilgardia 6~(11) (1932) 315--353.

\bibitem{Brody1945}
S.~Brody, Bioenergetics and growth, Reinhold Publishing Corp., New York (1945).

\bibitem{Cox2000}
C.~Cox, C.~Allen, Allen's Astrophysical Quantities, Springer Nature Book
  Archives Millennium, Springer (2000).

\bibitem{Thomas2008}
D.~Thomas, T.~Cote, L.~Lawhorne, S.~Levenson, L.~Rubenstein, D.~Smith,
  R.~Stefanacci, E.~Tangalos, J.~Morley, Understanding clinical dehydration and
  its treatment, Journal of the American Medical Directors Association 9 (2008)
  292--301.

\bibitem{Grandjean2003}
A.~C. Grandjean, K.~J. Reimers, M.~E. Buyckx, Hydration: Issues for the 21st
  century, Nutrition Reviews 61~(8) (2003) 261--271.

\bibitem{NRC2005}
P.~Water, S.~Intakes, Dietary reference intakes for water, potassium, sodium,
  chloride, and sulfate, National Academies Press (2005).

\bibitem{Tannenbaum2011}
S.~Tannenbaum, An investigation into equations for estimating water
  requirements and the development of new equations for predicting total water
  intake, PhD thesis, Florida International University (2011).

\bibitem{Moran1995}
D.~S. Moran, Y.~Shapiro, Y.~Epstein, R.~Burstein, L.~A. Stroschein, K.~B.
  Pandolf, Validation and adjustment of the mathematical prediction model for
  human rectal temperature responses to outdoor environmental conditions.,
  Ergonomics 38 5 (1995) 1011--8.

\bibitem{Shapiro1995}
Y.~Shapiro, D.~Moran, Y.~Epstein, L.~Stroschein, K.~B. Pandolf, Validation and
  adjustment of the mathematical prediction model for human sweat rate
  responses to outdoor environmental conditions, Ergonomics 38~(5) (1995)
  981--986.

\bibitem{Kellogg2019}
W.~Kellogg, Climate Change And Society: Consequences Of Increasing Atmospheric
  Carbon Dioxide, Westview Press (2019).

\bibitem{Seltenrich2015}
N.~Seltenrich, Between extremes: Health effects of heat and cold, Environmental
  health perspectives 123 (2015) A275--A279.

\bibitem{Coccarelli2017}
A.~Coccarelli, E.~Boileau, D.~Parthimos, P.~Nithiarasu, Modeling accidental
  hypothermia effects on a human body under different pathophysiological
  conditions, Medical \& Biological Engineering \& Computing (2017) 55.

\bibitem{Hollingsworth1969}
T.~Hollingsworth, Historical demography, The Sources of history, Cornell
  University Press (1969).

\bibitem{Kohn2007}
G.~Kohn, Encyclopedia of Plague and Pestilence: From Ancient Times to the
  Present, Facts on File Library of World History, Facts On File Incorporated
  (2007).

\bibitem{Lequeux2004}
J.~Lequeux, E.~Falgarone, C.~Ryter, The Interstellar Medium, Astronomy and
  Astrophysics Library, Springer Berlin Heidelberg (2004).

\bibitem{Zhang2019}
Y.~Zhang, S.~Mei, Y.~Li, Q.~Bi, Z.~Yang, Y.~Wang, X.~Chen, X.~Li, Study on
  bacterial diversities in stored water under different storage environments,
  {IOP} Conference Series: Earth and Environmental Science 267 (2019) 062001.

\bibitem{Duchesne2010}
S.~Duchesne, C.~Tressler, Environmental control and life support integration
  strategy for 6 crew operations, 40th International Conference on
  Environmental Systems (2010).

\bibitem{Proshkin2014}
V.~Proshkin, E.~Kurmazenko, Russian oxygen generation system "elektron-vm":
  hydrogen content in electrolytically produced oxygen for breathing by
  international space station crews, Aerospace and environmental medicine 48
  (2014) 65--8.

\bibitem{Shkedi2004}
B.~Shkedi, D.~Thompson, Iss eclss: 3 years of logistics for maintenance, SAE
  International (2004).

\bibitem{Bagdigian2005}
R.~Bagdigian, D.~Cloud, Status of the international space station regenerative
  eclss water recovery and oxygen generation systems, SAE International (2005).

\bibitem{Hale2011}
NASA, W.~Hale, H.~Lane, J.~Young, R.~Crippen, G.~Chapline, K.~Lulla, Wings in
  Orbit: Scientific and Engineering Legacies of the Space Shuttle, 1971-2010,
  NASA SP, U.S. Government Printing Office (2011).

\bibitem{Chan2002}
S.~Chan, X.~Chen, K.~Khor, An electrolyte model for ceramic oxygen generator
  and solid oxide fuel cell, Journal of Power Sources 111 (2002) 320–328.

\bibitem{Podgorodetskii2015}
G.~Podgorodetskii, Y.~Yusfin, A.~Sazhin, V.~Gorbunov, L.~Polulyakh, Production
  of generator gas from solid fuels, Steel in Translation 45 (2015) 395--402.

\bibitem{Chen2017}
L.~Chen, L.~Kong, J.~Bao, M.~Combs, H.~Nikolic, Z.~Fan, K.~Liu, Experimental
  evaluations of solid-fueled pressurized chemical looping combustion – the
  effects of pressure, solid fuel and iron-based oxygen carriers, Applied
  Energy 195 (2017) 1012--1022.

\bibitem{Liu2019}
J.-g. Liu, L.-z. Jin, N.~Gao, S.-n. Ou, S.~Wang, W.-x. Wang, A review on
  chemical oxygen supply technology within confined spaces: Challenges,
  strategies, and opportunities toward chemical oxygen generators (cogs),
  International Journal of Minerals, Metallurgy, and Materials 26 (2019)
  925--937.

\bibitem{Bockstahler2013}
K.~Bockstahler, J.~Lucas, J.~Witt, Design status of the advanced closed loop
  system acls for accommodation on the iss, 43rd International Conference on
  Environmental Systems (2013).

\bibitem{Putz2016}
D.~P{\"u}tz, C.~Olthoff, M.~Ewert, M.~Anderson, Assessment of the impacts of
  acls on the iss life support system using dynamic simulations in v-hab, 46th
  International Conference on Environmental Systems (2016).

\bibitem{Binot1994}
R.~Binot, C.~Tamponnet, C.~Lasseur, Biological life support for manned missions
  by esa, Advances in Space Research 14~(11) (1994) 71 -- 74.

\bibitem{Gros2003}
J.~Gros, L.~Poughon, C.~Lasseur, A.~Tikhomirov, Recycling efficiencies of c, h,
  o, n, s, and p elements in a biological life support system based on
  microorganisms and higher plants, Advances in space research : the official
  journal of the Committee on Space Research (COSPAR) 31 (2003) 195--9.

\bibitem{Tikhomirov2003}
A.~Tikhomirov, S.~Ushakova, N.~Manukovsky, G.~Lisovsky, Y.~Kudenko, V.~Kovalev,
  I.~Gribovskaya, L.~Tirranen, I.~Zolotukhin, J.~Gros, C.~Lasseur, Synthesis of
  biomass and utilization of plants wastes in a physical model of biological
  life-support system, Acta astronautica 53 (2003) 249--57.

\bibitem{Jones2019}
A.~Jones, China grew two leaves on the moon: The chang'e-4 spacecraft also
  carried potato seeds and fruit-fly eggs to the lunar far side - [news], IEEE
  Spectrum 56 (2019) 9--10.

\bibitem{Petersen1992}
J.~Petersen, A.~Haberstock, T.~Siccama, K.~Vogt, D.~Vogt, B.~Tusting, The
  making of biosphere 2, Ecological Restoration 10 (1992) 158--168.

\bibitem{Vernazza2016}
P.~{Vernazza}, P.~{Beck}, Composition of solar system small bodies, Cambridge
  University Press (2016).

\bibitem{Park2016}
R.~S. {Park}, A.~S. {Konopliv}, B.~G. {Bills}, N.~{Rambaux}, J.~C.
  {Castillo-Rogez}, C.~A. {Raymond}, A.~T. {Vaughan}, A.~I. {Ermakov}, M.~T.
  {Zuber}, R.~R. {Fu}, M.~J. {Toplis}, C.~T. {Russell}, A.~{Nathues},
  F.~{Preusker}, {A partially differentiated interior for (1) Ceres deduced
  from its gravity field and shape}, Nature 537~(7621) (2016) 515--517.

\bibitem{Platz2016}
T.~{Platz}, A.~{Nathues}, N.~{Schorghofer}, F.~{Preusker}, E.~{Mazarico}, S.~E.
  {Schr{\"o}der}, S.~{Byrne}, T.~{Kneissl}, N.~{Schmedemann}, J.~P. {Combe},
  M.~{Sch{\"a}fer}, G.~S. {Thangjam}, M.~{Hoffmann}, P.~{Gutierrez-Marques},
  M.~E. {Landis}, W.~{Dietrich}, J.~{Ripken}, K.~D. {Matz}, C.~T. {Russell},
  {Surface water-ice deposits in the northern shadowed regions of Ceres},
  Nature Astronomy 1 (2016) 0007.

\bibitem{Prettyman2017}
T.~H. {Prettyman}, N.~{Yamashita}, M.~J. {Toplis}, H.~Y. {McSween},
  N.~{Sch{\"o}rghofer}, S.~{Marchi}, W.~C. {Feldman}, J.~{Castillo-Rogez},
  O.~{Forni}, D.~J. {Lawrence}, E.~{Ammannito}, B.~L. {Ehlmann}, H.~G.
  {Sizemore}, S.~P. {Joy}, C.~A. {Polanskey}, M.~D. {Rayman}, C.~A. {Raymond},
  C.~T. {Russell}, {Extensive water ice within Ceres{\textquoteright} aqueously
  altered regolith: Evidence from nuclear spectroscopy}, Science 355~(6320)
  (2017) 55--59.

\bibitem{Correa2019}
J.~A. {Correa-Otto}, M.~F. {Calandra}, {Stability in the most external region
  of the Oort Cloud: evolution of the ejected comets}, Monthly Notices of the
  Royal Astronomical Society 490~(2) (2019) 2495--2506.

\bibitem{Engelhardt2017}
T.~{Engelhardt}, R.~{Jedicke}, P.~{Vere{\v{s}}}, A.~{Fitzsimmons},
  L.~{Denneau}, E.~{Beshore}, B.~{Meinke}, {An Observational Upper Limit on the
  Interstellar Number Density of Asteroids and Comets}, The Astronomical
  Journal 153~(3) (2017) 133.

\bibitem{Zdankiewicz1985}
E.~M. Zdankiewicz, D.~F. Price, Phase change water processing for space
  station, SAE Transactions 94 (1985) 286--295.

\bibitem{Ronsch2016}
S.~R{\"o}nsch, J.~Schneider, S.~Matthischke, M.~Schlüter, M.~G{\"o}tz,
  J.~Lefebvre, P.~Prabhakaran, S.~Bajohr, Review on methanation – from
  fundamentals to current projects, Fuel 166 (2016) 276 -- 296.

\bibitem{Luo2017}
Y.~Luo, X.-Y. Wu, S.~Yixiang, A.~Ghoniem, N.~Cai, Exergy efficiency analysis of
  a power-to-methane system coupling water electrolysis and sabatier reaction,
  ECS Transactions 78 (2017) 2965--2973.

\bibitem{Messerschmid2013}
E.~Messerschmid, T.~Freyer, R.~Bertrand, Space Stations: Systems and
  Utilization, Springer Berlin Heidelberg (2013).

\bibitem{Anderson2015}
M.~S. Anderson, M.~K. Ewert, J.~F. Keener, S.~A. Wagner, Life support baseline
  values and assumptions document, National Aeronautics and Space
  Administration Technical Report JSC-CN-32628 (2012).

\bibitem{Langston2012}
L.~Langston, Generic groundrules, requirements, and constraints part 1:
  Strategic and tactical planning, National Aeronautics and Space
  Administration, International Space Station Program SPP 50261-01 (2012).

\bibitem{Czupalla2007}
M.~Czupalla, T.~Dirlich, S.~I. Bartsev, An approach to lss optimization based
  on equivalent system mass, system stability and mission success, Tech. rep.,
  SAE Technical Paper (2007).

\bibitem{Plotner2013}
P.~Pl{\"o}tner, M.~Czupalla, A.~Zhukov, Closed environment module –
  modularization and extension of the virtual habitat, Advances in Space
  Research 52~(12) (2013) 2180 -- 2191.

\bibitem{Czupalla2015}
M.~Czupalla, A.~Zhukov, J.~Schnaitmann, C.~Olthoff, M.~Deiml, P.~Pl{\"o}tner,
  U.~Walter, The virtual habitat – a tool for dynamic life support system
  simulations, Advances in Space Research 55~(11) (2015) 2683 -- 2707.

\end{thebibliography}

\end{document}